\documentclass[10pt,superscriptaddress,prc,aps,twocolumn,showpacs,fleqn]{revtex4-1}
\usepackage{amsmath,amssymb}
\usepackage{graphicx}
\usepackage[usenames]{color}

\newcommand{\Mainz}[1]
{\affiliation{Institut f\"ur Kernphysik, Johannes Gutenberg-Universit\"at Mainz, D-55099
Mainz,Germany}}

\begin{document}

\title{\boldmath Regge phenomenology in $\pi^0$ and $\eta$ photoproduction}

\author{V.~L.~Kashevarov}\thanks{kashev@kph.uni-mainz.de}\Mainz \\
\author{M.~Ostrick}\thanks{ostrick@kph.uni-mainz.de}\Mainz \\
\author{L.~Tiator}\thanks{tiator@uni-mainz.de}\Mainz \\

\date{\today}

\begin{abstract}

The $\gamma N \to \pi^0 N$ and $\gamma N \to \eta N$ reactions at
photon beam energies above 4 GeV are investigated within Regge models.
The models include $t$-channel exchanges of vector ($\rho$ and $\omega$)
and axial vector ($b_1$ and $h_1$) mesons. Moreover, Regge cuts of
$\rho{\mathbb P}$, $\rho f_2$, $\omega{\mathbb P}$, and $\omega f_2$
are investigated. A good description of differential cross sections
and polarization observables at photon beam energies from 4 to 15 GeV
can be achieved.
\end{abstract}

\maketitle

\section{Introduction}

Meson photo- and electroproduction processes are closely related to the
long-range structure and dynamics of hadrons.
The phenomenology of these reactions changes at center of mass energies of about
$W\approx 3$~GeV, roughly separating resonance and continuum regions.

Below $W\approx 3 $~GeV, which corresponds to photon beam energies
below $E_{\gamma} \approx 4$~GeV, the reaction dynamics is
characterized by the excitation of individual s-channel baryon
resonances with definite quantum numbers on top of smooth,
non-resonant background. Within the last two decades, new data on
photoinduced meson production has become the major source of
information for baryon spectroscopy. At the electron accelerator
labs ELSA, JLab and MAMI  extensive developments in beam and target
polarization techniques have been undertaken and an enormous amount
of data with different types of polarization has been obtained,
especially for $\pi$, $\eta$ and $K$ photoproduction
\cite{Crede:2013sze}.
Above this resonance region, at $W \gtrsim  3$~GeV, the reaction
dynamics changes and can be described most effectively by particle
(reggeon) exchanges in the crossed $t$-channel \cite{Irving:1977ea}.
Experimental data on $\pi$ and $\eta$ photoproduction in this
high-energy region were mainly measured in the 1970s at
DESY~\cite{DESY:1970, DESY:1973, DESY:1968} and
SLAC~\cite{SLAC:1971}, but only a limited amount of target and
recoil polarization data is available. Only recently, the new GlueX
experiment in Hall-D at JLab started data taking and first results
on differential cross sections with a linearly polarized photon beam
at $E_{\gamma}$= 8.7 GeV were already obtained \cite{GlueX:2017}.

The resonance and the continuum regions are of course not
independent from each other but analytically connected via
dispersion
relations~\cite{Aznauryan:2002gd,Aznauryan:2003zg,Pasquini:2006yi,Pasquini:2007fw}
or finite energy sum
rules~\cite{Dolen:1967jr,Mathieu:2015-2,Nys:2017}. The motivation
for this study is therefore twofold. Firstly, with view to new
results on unpolarized cross sections and photon beam asymmetries
expected from GlueX in the next years, we want to obtain a deeper
understanding of the high-energy Regge phenomenology.

Secondly, we consider a good description of the high-energy data as an important
prerequisite for a  high-quality baryon resonance analysis at lower energies. In particular in
$\eta$, $\eta'$ and $K$ photoproduction a good knowledge about Regge contributions to
non-resonant background amplitudes is crucial for a reliable extraction of resonance
parameters.

The main features of our models are Regge trajectories from $\omega$
and $\rho$ vector mesons and Regge cuts arising from the exchange of
two Reggeons. We compare different approaches to available
high-energy data for $\pi^0$ and $\eta$ photoproduction at lab
energies above 4~GeV. We show that in particular polarization
observables, as photon beam and target asymmetries or recoil
polarization, are crucial to distinguish between the different
models.

This paper is organized as follows. In section II we briefly
introduce kinematics, polarization observables and photoproduction
amplitudes. In section III we compare different Regge approaches
with Regge poles and Regge cuts and discuss the various
trajectories. In section IV we compare  different models to
high-energy data of $\pi^0$ and $\eta$ photoproduction for
unpolarized cross sections and polarization observables.

\section{Kinematics, observables and amplitudes}

\subsection{Kinematics}

Let us first define the kinematics of $\pi$ and  $\eta$
photoproduction reactions on a nucleon,
\begin{equation}
\gamma (k) + N(p_i)\to\pi/\eta(q) + N'(p_f)\ ,
\end{equation}
where the variables in brackets denote the four-momenta of the
participating particles. The familiar Mandelstam variables are
\begin{equation}
s=(p_i+k)^2,\quad t=(q-k)^2,\quad u=(p_i-q)^2\,,
\end{equation}
where the sum of the Mandelstam variables is given by the sum of the
external masses.
The crossing symmetrical variable $\nu$
is related to the photon lab energy $E_\gamma^{lab}$ by
\begin{equation}
 \nu=\frac{(s-u)}{4M_N} = E_\gamma^{lab} + \frac{t-\mu^2}{4M_N}\,,
\end{equation}
where $M_N$ and $\mu$ are nucleon and meson masses ($\pi$ or $\eta$),
respectively.

\subsection{Observables}

In photoproduction of pseudoscalar mesons a total of 16 polarization
observables can be measured, which include the unpolarized cross
section, three single-polarization and 12 double-polarization
observables. By considering only beam and target polarization, the
cross section depends on 8 observables, which can be separated by
circular, $P_{\odot}$, and linear, $P_T$, photon beam polarization
and the three components $P_x,P_y,P_z$ of the target polarization vector:
\begin{eqnarray}
\frac{d \sigma}{d \Omega} &=& \sigma_0 \left\{ 1 - P_T \Sigma \cos 2 \varphi \right. \nonumber \\
                                & & + P_x \left( - P_T H \sin 2 \varphi + P_{\odot} F \right) \nonumber \\
                                & & - P_y \left( - T + P_T P \cos 2 \varphi \right) \nonumber \\
                                & & \left. - P_z \left( - P_T G \sin 2 \varphi + P_{\odot} E \right)
                                \right\}\,.
\end{eqnarray}
The $z$-axis is pointing into the direction of the incoming photon. The $\hat y$ direction is
perpendicular to the reaction plane, $\hat y = \hat z \times \hat q$,  defined by the incoming
photon and the direction of the outgoing meson $\hat q$. The $x$-axis is given by
\begin{figure}
\begin{center}
\resizebox{0.35\textwidth}{!}{\includegraphics{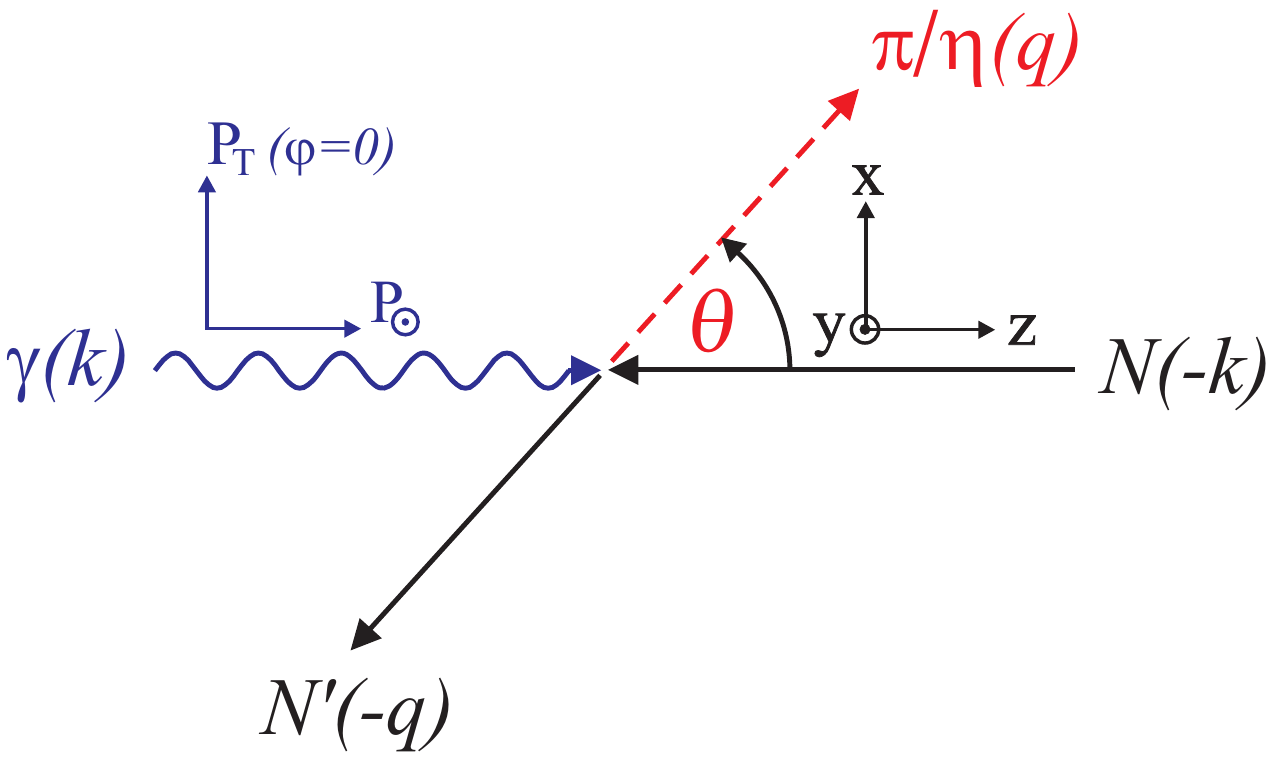}}
\caption{Kinematics for $\pi^0$ or $\eta$ photoproduction and frames
for beam and target polarization.} \label{fig:kin}
\end{center}
\end{figure}
$\hat x = \hat y \times \hat z$. The orientation of the linear
polarization vector of the photon beam relative to the production
plane is given by the angle $\varphi$, see Fig.~\ref{fig:kin}.
Expressions of the polarization observables in terms of amplitudes
are given in the appendix.

\subsection{\boldmath Invariant amplitudes and fixed-$t$ dispersion relations}

The electromagnetic current for pseudoscalar meson photoproduction
can be expressed in terms of four invariant amplitudes
$A_i(\nu,t)$~\cite{CGLN:1957},
\begin{eqnarray}
J^\mu = \sum_{i=1}^4 A_i(\nu,t)\, M^\mu_i,
\end{eqnarray}
with the gauge-invariant four-vectors $M^\mu_i$ given by
\begin{eqnarray}
M^\mu_1&=&
-\frac{1}{2}i\gamma_5\left(\gamma^\mu\sl{k}-\sl{k}\gamma^\mu\right)\, ,
\nonumber\\
M^\mu_2&=&2i\gamma_5\left(P^\mu\, k\cdot(q-\frac{1}{2}k)-
(q-\frac{1}{2}k)^\mu\,k\cdot P\right)\, ,\nonumber\\
M^\mu_3&=&-i\gamma_5\left(\gamma^\mu\, k\cdot q
-\sl{k}q^\mu\right)\, ,\nonumber\\\
M^\mu_4&=&-2i\gamma_5\left(\gamma^\mu\, k\cdot P
-\sl{k}P^\mu\right)-2M_N \, M^\mu_1\, ,
 \label{eq:tensor}
\end{eqnarray}
where $P^\mu=(p_i^\mu+p_f^\mu)/2$.%

The invariant amplitudes $A_i(\nu,t)$ have definite crossing symmetry and
satisfy the following dispersion relations at fixed~$t$:
\begin{equation}
{\rm Re} A_i(\nu,t) = A_i^{pole}(\nu,t) +\frac{2}{\pi}\;{\cal
P}\hspace{-6pt}\int_{\nu_{thr}}^{\infty}{\rm d}\nu'\;
\frac{\nu'\,{\rm Im}A_i(\nu',t)}{\nu'^2-\nu^2}\,, \label{eq:dr1}
\end{equation}
for the crossing-even amplitudes, $A_{1,2,4}$,  and
\begin{equation}
{\rm Re} A_3(\nu,t) = A_3^{pole}(\nu,t) +\frac{2\nu}{\pi}\;{\cal
P}\hspace{-6pt}\int_{\nu_{thr}}^{\infty}{\rm d}\nu'\; \frac{{\rm
Im}A_3(\nu',t)}{\nu'^2-\nu^2}\, \label{eq:dr2}
\end{equation}
for the crossing-odd amplitude $A_{3}$~\cite{Pasquini:2006yi}.

\section{\boldmath $t$-channel exchanges}

\subsection{\boldmath Vector and axial-vector poles in the $t$ channel}

The amplitudes of pseudoscalar meson photoproduction typically
contain contributions from nucleon resonance excitations and a non-resonant
background from Born terms and $t$-channel meson
exchanges. In the current approach we want to consider only amplitudes at
high energies beyond the nucleon resonance region. Furthermore,
we neglect Born terms, which are practically zero for $\eta$
photoproduction~\cite{MAID:2003}. Also in $\pi^0$ photoproduction
they only play a minor role at forward angles.

\begin{figure*}
\begin{center}
\resizebox{0.8\textwidth}{!}{\includegraphics{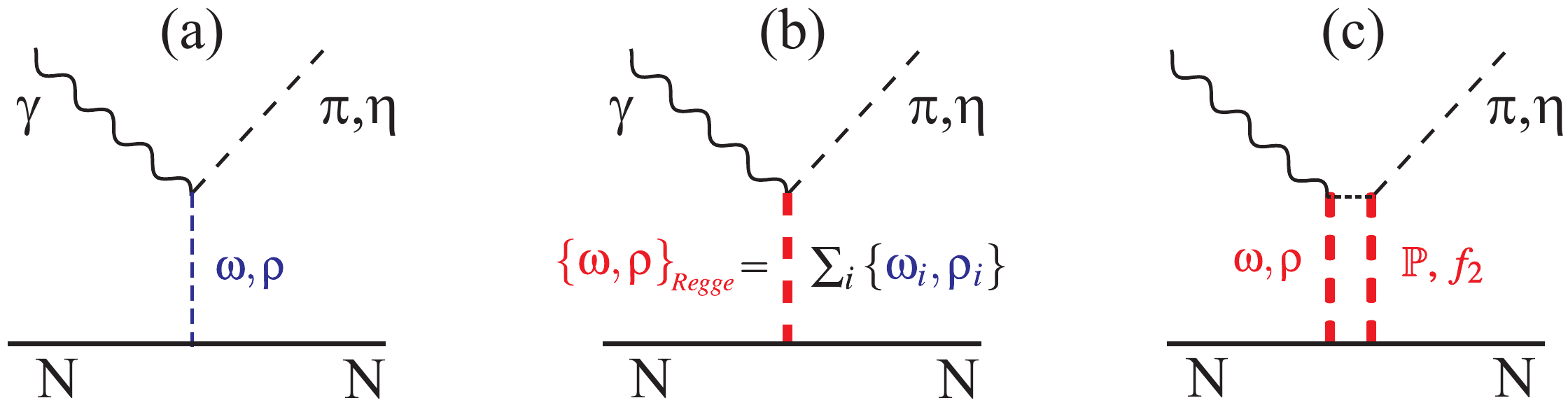}}
\caption{$t$-channel contributions to $\eta$ photoproduction from
single poles (a), Regge poles (b), and Regge cuts (c). An example
for $\rho$ and $\omega$ meson exchange and $\mathbb P$ and $f_2$
mesons for rescattering of two Reggeons.} \label{fig:regge}
\end{center}
\end{figure*}

We concentrate on $t$-channel contributions and will
firstly consider the exchange of vector and axial vector mesons in
terms of single pole Feynman diagrams,
see Fig.~\ref{fig:regge}(a) as an example for $\rho$ and $\omega$ meson exchange.

Expressed in terms of invariant amplitudes $A_i$, these $t$-channel
Feynman diagrams obtain the simple form
%
\begin{eqnarray}
A_1(t) & = & \frac{e\,\lambda_V\,g_V^{\mathfrak{t}}}{2\mu M_N}\;
\frac{t}{t-M_V^2}\, ,    \label{Eq:A1}\\
A_2'(t) & = & - \frac{e\,\lambda_A\,g_A^{\mathfrak{t}}}{2\mu M_N} \;
\frac{t}{t-M_A^2}\, ,   \label{Eq:A2}\\
A_3(t)& = & \frac{e\,\lambda_A\,g_A^v}{\mu}\; \frac{1}{t-M_A^2}\, ,   \label{Eq:A3}\\
A_4(t) & = & \frac{-e\,\lambda_V\,g_V^v}{\mu}\;
\frac{1}{t-M_V^2}\,,\label{Eq:A4}
\end{eqnarray}
%
where $\lambda_{V(A)}$ denotes the electromagnetic coupling of the
vector ($V$) or axial ($A$) vector mesons with masses $M_{V(A)}$.
The constants $g_{V(A)}^{v(\mathfrak{t})}$ denote their vector $(v)$
or tensor $(\mathfrak{t})$ couplings to the nucleon. In order to
separate the vector and tensor contributions from individual mesons,
we followed Ref.~\cite{Nys:2017} and introduced the amplitude
\begin{equation}\label{Eq:A2p}
A_2'(t)=A_1(t)+t\,A_2(t)\, ,
\end{equation}
which has only contributions from the tensor coupling of an axial vector exchange.

\begin{table*}     
\caption{Isospin $I$, $G$-parity, spin $J$, parity $P$, and charge
conjugation $C$ quantum numbers for pseudoscalar, vector and axial
vector mesons. \label{tab:mesons}}
\begin{tabular}{|c|ccc|ccccccc|}
\hline
         & $\gamma$ & $\pi^0$  & $\eta$   & $\rho(770)$& $\omega(782)$&
$\phi(1020)$& $b_1(1235)$& $h_1(1170)$& $a_1(1260)$& $f_1(1285)$\\
\hline
$I^G$    & $0,1$    & $1^-$    & $0^+$    & $1^+$      & $0^-$
& $0^-$      & $1^+$      & $0^-$      & $1^-$      & $0^+$ \\
$J^{PC}$ & $1^{--}$ & $0^{-+}$ & $0^{-+}$ & $1^{--}$   & $1^{--}$
& $1^{--}$   & $1^{+-}$   & $1^{+-}$   & $1^{++}$   & $1^{++}$ \\
\hline
\end{tabular}
\end{table*}
There are three vector mesons $\rho$, $\omega$, $\phi$  and four
axial vector mesons $b_1$, $h_1$, $a_1$, $f_1$, that could be used in
our approach. The details on the quantum numbers are listed in
Table~\ref{tab:mesons}. For the nucleon vertex, the axial-vector
coupling $\gamma^\mu\gamma_5$ is $C$-even and the pseudo-tensor
coupling $\sigma^{\mu\nu}\gamma_5$ is $C$-odd~\cite{Kaskulov:2010}.
Therefore, due to charge conjugation conservation, the $C$-odd $b_1$
and $h_1$ mesons couple to the nucleon via the tensor coupling only
and can contribute to the $A_2$ ($A_2^\prime$) amplitude (see
equations (\ref{Eq:A2}) and (\ref{Eq:A2p})), whereas $C$-even $a_1$
and $f_1$ mesons via the vector coupling only and, in principal, can
contribute to the $A_3$ amplitude. However, the quantum numbers
$I^G$ should be equal to $0^-$ or $1^+$ for $\pi^0$ and $\eta$
photoproduction on the nucleon.
Consequently, $a_1(I^G=1^-)$ and $f_1(I^G=0^+)$ are
excluded in our case. The $a_1$ is a good candidate for charged-pion
photoproduction and $f_1$ for the $\gamma p \to \rho^0 p$
channel~\cite{Pasquini:2006yi}. Therefore, there is no candidate left among
vector and axial vector mesons which could contribute to $A_3$.

The $\phi$ meson could in principle contribute to $A_1$ and $A_4$.
However, being practically a pure strange quark-antiquark state,
a very small coupling to the nucleon is expected and it is commonly
neglected in $\pi^0$ and $\eta$ photoproduction.

The invariant amplitudes (\ref{Eq:A1})-(\ref{Eq:A4}) contain
only the product of electromagnetic and hadronic coupling constants.
We have fixed one of them and determined the second one by the
fit. In general, the values for the strong coupling constants
$g^{v}$ and $g^{\mathfrak{t}}$ are not well known, especially
for the axial vector mesons. Results for these constants from
different analyses and models are summarized in Ref.~\cite{Yu:2011}, Table IV.
Therefore, in our present work, we fix the electromagnetic couplings $\lambda_{V(A)}$.
For $\pi^0$ and
$\eta$ photoproduction they can be determined from the radiative widths
$\Gamma_{V(A)}$ of the decays $V(A) \rightarrow \pi^0\gamma$ and
$V(A) \rightarrow \eta\gamma$, respectively,
\begin{eqnarray}  \label{Eq:Gamma}
 \Gamma_{V(A)}
 &=& \frac{\alpha(M_{V(A)}^2-\mu^2)^3}{24\,M_{V(A)}^3\,\mu^2}\,
     \lambda_{V(A)}^2 \,,
\end{eqnarray}
where $\alpha$ is the fine-structure constant. For
$\lambda_{V\pi^0\gamma}$ we used the decay widths $\Gamma_{\rho
\rightarrow \pi^0\gamma} = 91.0\ \mathrm{keV}$ and $\Gamma_{\omega
\rightarrow \pi^0\gamma} = 703.0\ \mathrm{keV}$.
In case of the $\eta$ meson, we determined $\lambda_{V\eta\gamma}$
from $\Gamma_{\rho \rightarrow
\eta\gamma} = 50.6\ \mathrm{keV}$ and $\Gamma_{\omega \rightarrow
\eta\gamma} = 3.9\ \mathrm{keV}$~\cite{PDG2016}.
For the $b_1$ meson only the
electromagnetic width for the charged decay
$\Gamma_{b_1 \rightarrow \pi^{\pm}\gamma} = 227\ \mathrm{keV}$ is known ~\cite{PDG2016}.
We use this value to calculate $\lambda_{b_1}$ for the neutral decay as well,
because chiral unitary models predict practically the same
electromagnetic couplings of the $b_1$ meson for both charged and
neutral pion decays~\cite{Chiral:2008}. Unfortunately there are no
data for the decay $b_1 \to \eta \gamma$. In this case, we arbitrarily fixed
$\lambda_{\eta\gamma} = 0.1$ which is close to the value obtained
for the $\pi \gamma$ decay.
All electromagnetic coupling constants for the $\rho$, $\omega$ and
$b_1$ mesons used in the present work are listed in
Table~\ref{tab:Fit1fix}. For the contribution of the $h_1$ meson
we follow Ref.~\cite{Nys:2017} that
suggests a fraction of $2/3$ of the $b_1$ contribution.

\subsection{\boldmath Regge trajectories and $t$-channel Regge amplitudes}

\begin{figure*}    
\begin{center}
\resizebox{0.8\textwidth}{!}{\includegraphics{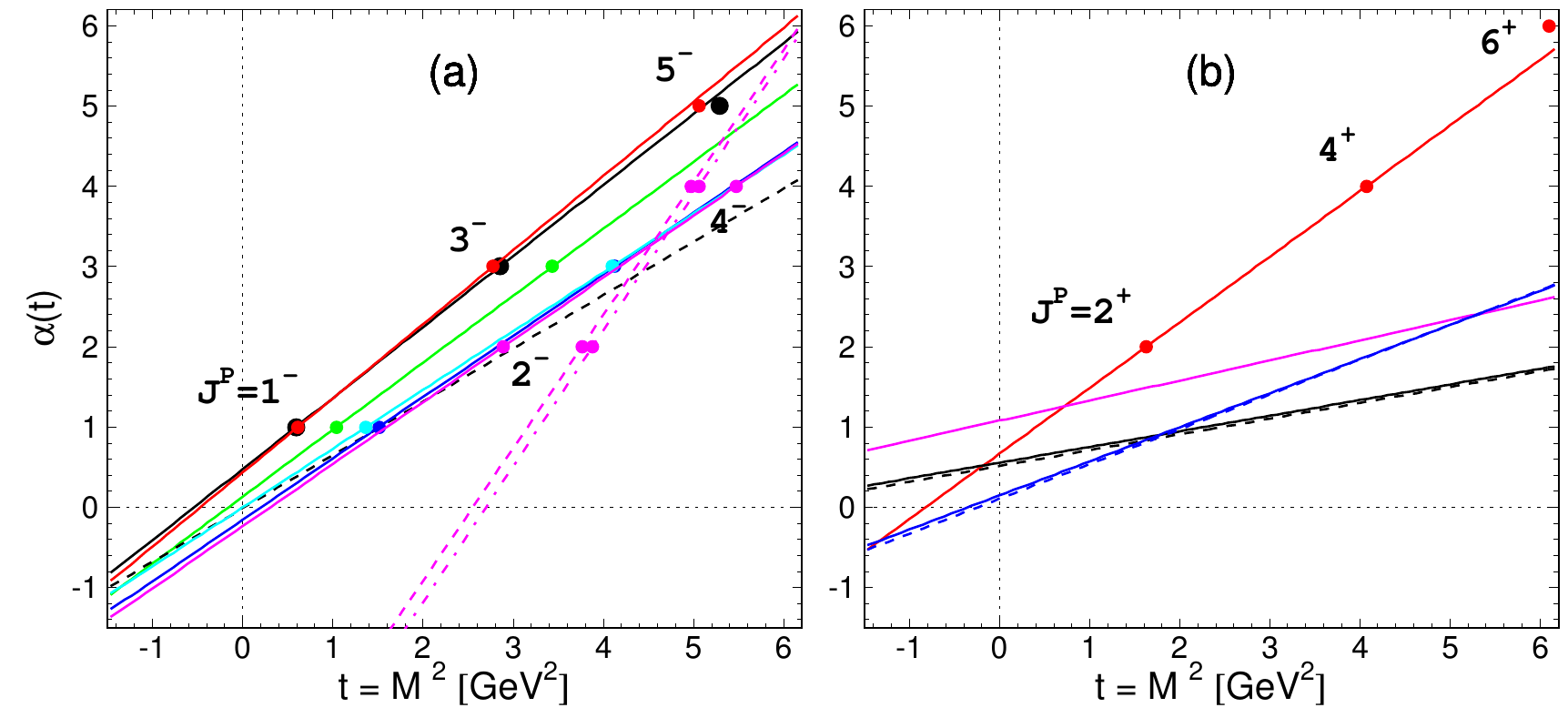}} \caption{Regge
trajectories: (a) $\rho$ black, $\omega$ red, $\phi$ blue, $b_1$ and
$h_1$ green, $\rho_2$ and $\omega_2$ magenta; dashed and dash-dotted
magenta lines are $\rho_2$ and $\omega_2$ of
Ref.~\cite{Anisovich:2002-1,Anisovich:2002-2}; (b) $f_2$ red,
$\mathbb P$ magenta, $\rho f_2$ black solid, $\omega f_2$ blue
dashed, $\rho \mathbb P$ black solid, $\omega \mathbb P$ black
dashed. } \label{fig:traj}
\end{center}
\end{figure*}

Mesons fall into linear trajectories when their spin is plotted
against the squared meson masses (Chew-Frautchi-Plot). These Regge
trajectories are usually parameterized as
\begin{equation}
\alpha(t) = \alpha_{0} + \alpha^\prime\,t,
\end{equation}
see e.g. Ref.~\cite{Collins:1977}. Examples of such trajectories are
shown in Fig.~\ref{fig:traj}(a).

It can be assumed that in photoproduction reactions not only single
mesons but whole Regge trajectories are exchanged in the t-channel
as illustrated in  Fig.~\ref{fig:regge}(b). In our models we include
the $\rho$, $\omega$, $\phi$, and $b_1$ trajectories shown in
Fig.~\ref{fig:traj}(a). The trajectory for the $h_1$ is assumed to
be the same as for the $b_1$.
Furthermore, trajectories for tensor mesons $\rho_2$ and $\omega_2$
are shown in the same plot. These mesons, assuming the same
masses for both, were predicted in a relativized quark
model~\cite{Godfrey:1985} for two states: $J^{PC}=2^{--}$ with mass
of $1.7$~GeV and $J^{PC}=4^{--}$ with mass of $2.34$~GeV. The
trajectory drawn through these two points is shown by the magenta
line. According to their quantum numbers, the $\rho_2$ and
$\omega_2$ could be good candidates for the $A_3$ amplitude in
$\pi^0$ and $\eta$ photoproduction. However, there is no clear
experimental evidence for the existence of these states. They were
found in a partial wave analysis of
Refs.~\cite{Anisovich:2002-1,Anisovich:2002-2} and result in much
steeper trajectories, that are shown in Fig.~\ref{fig:traj}(a) by
the dashed magenta line for the $\rho_2$ and dash-dotted magenta
line for the $\omega_2$.

Technically, the t-channel exchange of Regge trajectories is done by replacing
the single meson propagator by the following expression
\begin{equation}
\frac{1}{t-M^2} \Rightarrow
\left(\frac{s}{s_0}\right)^{\alpha(t)-1}\;
\frac{\pi\,\alpha^\prime}{\mbox{sin}[\pi\alpha(t)]}\;
\frac{{\cal S} + e^{-i\pi\alpha(t)}}{2}\;
\frac{1}{\Gamma(\alpha(t))}\,,
\label{eq:Regge-1}
\end{equation}
where $M$ is the mass of the Reggeon, $\cal S$ is the signature of
the Regge trajectory, and $s_0$ is a mass scale factor, commonly set
to 1~$GeV^2$. The Gamma function $\Gamma(\alpha(t))$ is introduced
to suppress additional poles of the propagator.
\begin{figure}[bh]
\begin{center}
\resizebox{0.48\textwidth}{!}{\includegraphics{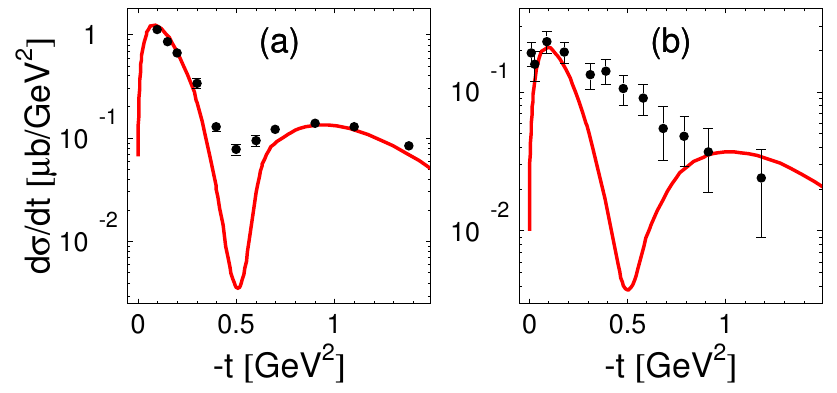}} \caption{The
differential cross sections of $\gamma p \to \pi^0 p$ (a) and
$\gamma p \to \eta p$ (b) reactions at $E_{\gamma}=6$~GeV.
Experimental data are from Ref.~\cite{SLAC:1971} (a) and
Ref.~\cite{DESY:1970} (b). The solid line is a calculation with
$\rho$ and $\omega$ exchange in the $t$ channel.} \label{fig:dip}
\end{center}
\end{figure}
The signature $\cal S$ is determined as ${\cal S}= (-1)^J$ for
bosons and  ${\cal S}= (-1)^{J+1/2}$ for fermions. So ${\cal S}= -1$
for the vector and axial-vector mesons, and ${\cal S}= +1$ for
tensor mesons. If ${\cal S}= -1$ and $\alpha(t)= 0$, then both, real
and imaginary parts, vanish. This results in a characteristic dip of
differential cross sections of $\gamma p \to \pi^0 p$ and $\gamma p
\to \eta p$ reactions at $t\approx -0.5$~GeV$^2$, which is not
observed in experimental data, see Fig.~\ref{fig:dip}.

To avoid problems with the dip at $\alpha(t) = 0$, different
approaches have been developed, see for example Ref.~\cite{Nys:2017,
MAID:2003, Sibirtsev:2016, Goldstein:1973, Barker:1978, DoKa:2016,
Mathieu:2015}. Here we focus on two of them, which are described in
the following subsections.

\subsection{Regge cuts}

Regge cuts were firstly considered in the early work of
Refs.~\cite{Landshoff:1972, Goldstein:1973, Barker:1978}, where their important role
was shown to fill in the dip in the differential cross sections of
$\pi^0$ and $\eta$ photoproduction. A full discussion of Regge cuts
can be found in Ref.~\cite{Donnachie:2002}. In 2016 Donnachie and
Kalashnikova~\cite{DoKa:2016} revisited the Regge cuts and developed
a new approach, where in addition to Regge trajectories of $\rho$,
$\omega$, and $b_1$ exchange, also Regge cuts from rescattering
$\rho{\mathbb P}$, $\rho f_2$ and $\omega{\mathbb P}$, $\omega f_2$
were added, where ${\mathbb P}$ is the Pomeron with quantum numbers
of the vacuum $0^+(0^{++})$ and $f_2$ is a tensor meson with quantum
numbers $0^+(2^{++})$. These Regge cuts can be considered as
contracted box diagrams, where two particles are exchanged, see
Fig.~\ref{fig:traj}(c).

The exchange of two Reggeons with linear trajectories
\begin{equation}\label{regge_traj}
\alpha_i(t) = \alpha_i(0)+\alpha_i^\prime t,\quad  i=1,2
\end{equation}
yields a cut with a linear trajectory
$\alpha_c(t)$~\cite{Landshoff:1972}
\begin{equation}\label{regge_cut1}
\alpha_c(t) = \alpha_c(0)+\alpha_c^\prime\,t\,,
\end{equation}
where
\begin{eqnarray}\label{regge_cut2}
\alpha_c(0) &=& \alpha_1(0)+\alpha_2(0)-1\,, \nonumber \\
\alpha_c^\prime &=& \frac{\alpha_1^\prime\alpha_2^\prime}
{\alpha_1^\prime + \alpha_2^\prime}\,.
\end{eqnarray}

The trajectories for $f_2$ and ${\mathbb P}$ are shown in
Fig~\ref{fig:traj}(b) together with four cut trajectories
$\rho{\mathbb P}$, $\omega{\mathbb P}$ (black solid and dashed
lines) and $\rho f_2$, $\omega f_2$ (blue solid and dashed lines)
calculated by
Eqs.~(\ref{regge_traj},\ref{regge_cut1},\ref{regge_cut2}).
Parameters of the Reggeon and cut trajectories used in the present
work are collected in Table~\ref{tab:traject}.
\begin{table}[ht]
\caption{\label{tab:traject}The Reggeon and cut trajectories used in the present work.}
\begin{tabular}{|c|c|}
\hline
Reggeon or cut & $\alpha(t)$     \\
\hline
$\rho$                 & $ 0.477 + 0.885 \,t$ \\
$\omega$               & $ 0.434 + 0.923 \,t$ \\
$b_1$, $h_1$,          & $-0.013 + 0.664 \,t$ \\
$\rho_2$, $\omega_2$   & $-0.235 + 0.774 \,t$ \\
$f_2$,                 & $ 0.671 + 0.817 \,t$ \\
$\mathbb P$            & $ 1.08  + 0.25 \,t $ \\
 \hline
$\rho f_2$             & $ 0.148 + 0.425 \,t$ \\
$\omega f_2$           & $ 0.106 + 0.436 \,t$ \\
$\rho \mathbb P$       & $ 0.557 + 0.195 \,t$ \\
$\omega \mathbb P$     & $ 0.514 + 0.197 \,t$ \\
 \hline
\end{tabular}
\end{table}

All four Regge cuts can
contribute to vector and axial vector exchanges and can be written
in the following form
\begin{equation}
D_{cut} =\left(\frac{s}{s_0}\right)^{\alpha_c(t)-1}\;
e^{-i\pi\alpha_c(t)/2}\; e^{d_c t} \,.
\end{equation}

In total, the vector meson propagators are replaced by
\begin{equation}
D_V = {D}_{V} + c_{V\mathbb P}\,{D}_{V\mathbb P} +
c_{Vf_2}\,D_{Vf_2},\; V=\rho,\omega
\end{equation}
and the axial vector meson propagators are replaced by
\begin{equation}
D_A = {D}_{A} + \sum_{V=\rho,\omega} ({\tilde
c}_{V\mathbb P}\,{D}_{V\mathbb P}
   + {\tilde c}_{Vf_2}\,D_{Vf_2}),\; A=b_1, h_1\,,
\end{equation}
where the coefficients $c_{V\mathbb P},c_{Vf_2}$ are for natural parity
cuts and ${\tilde c}_{V\mathbb P},{\tilde c}_{Vf_2}$ for un-natural
parity cuts and are obtained by a fit to the data.

\begin{table*}      
\caption{\label{tab:VMInv}Vector and axial vector contributions to
invariant amplitudes.}
\begin{tabular}{|c|c|c|c|c|}
\hline
 $\eta$  & $J^P$           & Dirac coupling & Invariant amplitudes & Reggeons and cuts\\
\hline
 natural &$1^-,3^-,\ldots$ & $g_V^{v}\gamma^\mu$                 &$A_4$ & $\rho, \omega,
 \rho{\mathbb P}, \omega{\mathbb P}, \rho f_2, \omega f_2$ \\
 natural &$1^-,3^-,\ldots$ & $g_V^{\mathfrak{t}}\sigma^{\mu\nu}$ &$A_1$ & $\rho, \omega,
 \rho{\mathbb P}, \omega{\mathbb P}, \rho f_2, \omega f_2$ \\
\hline
 un-natural &$2^-,4^-\ldots$ & $g_A^{v}\gamma^\mu\gamma_5$      &$A_3$ & $\rho_2, \omega_2,
 \rho f_2, \omega f_2$ \\
 un-natural &$1^+,3^+,\ldots$ & $g_A^{\mathfrak{t}}\sigma^{\mu\nu}\gamma_5$ &$A_2'$ & $b_1, h_1,
 \rho f_2, \omega f_2$ \\
\hline
 \end{tabular}
 \end{table*}

In detail, the invariant amplitudes will be changed in the
following way
\begin{eqnarray}\label{Eq:cuts}
\begin{split}
\lambda_\rho\,g_\rho^{v,\mathfrak{t}}\;\frac{1}{t-M_\rho^2} &\rightarrow \lambda_\rho\,g_\rho^{v,\mathfrak{t}}\\
&\hspace{-1.8cm}[D_\rho(s,t) + c_{\rho \mathbb P}\,D_{\rho \mathbb P}(s,t) + c_{\rho f}\,D_{\rho f}(s,t)] \,,\\
\lambda_\omega\,g_\omega^{v,\mathfrak{t}}\;\frac{1}{t-M_\omega^2}
&\rightarrow
\lambda_\omega\,g_\omega^{v,\mathfrak{t}}\\
&\hspace{-1.8cm}[D_\omega(s,t) + c_{\omega \mathbb P}\,D_{\omega \mathbb P}(s,t) + c_{\omega f}\,D_{\omega f}(s,t)]\,,\\
\lambda_{b_1}\,g_{b_1}^{\mathfrak{t}}\;\frac{1}{t-M_{b_1}^2}
&\rightarrow \lambda_{b_1}\,g_{b_1}^{\mathfrak{t}}
D_{b_1}(s,t)\\
&\hspace{-1.8cm}+\lambda_\rho    \,g_\rho^{\mathfrak{t}}\,[{\tilde
c}_{\rho \mathbb P}  \,D_{\rho \mathbb P}(s,t)
 + {\tilde c}_{\rho f_2}  \,D_{\rho f_2}(s,t)] \\
&\hspace{-1.8cm}+\,\lambda_\omega\,g_\omega^{\mathfrak{t}}\,[{\tilde
c}_{\omega \mathbb P}\,D_{\omega \mathbb P}(s,t)
 + {\tilde c}_{\omega f_2}\,D_{\omega f_2}(s,t)]\,.
\end{split}
\end{eqnarray}

In practical calculations, it turns out that the axial vector Regge
pole contributions, proportional to $D_A$, can be neglected, but the
axial vector Regge cuts arising from $\rho$ and $\omega$ together
with $\mathbb P$ and $f_2$ are very important, in particular for
polarization observables, as the photon beam asymmetry $\Sigma$.

The Regge cuts also allow us to describe a long standing problem of
suitable candidates for an $A_3$ amplitude: $\rho f_2$ and $\omega
f_2$ satisfy all conservation law requirements. In
Table~\ref{tab:VMInv} details of the invariant amplitude structure
of the $t$-channel exchanges are given. Here, $\eta$ is a
naturality, determined as $\eta = P (-1)^J$. For the $\rho{\mathbb
P}$ and $\omega{\mathbb P}$ cuts, $\eta = +1$ and these cuts do not
contribute to the $A_3$ amplitude. Therefore, we set the
coefficients ${\tilde c}_{\rho \mathbb P}$ and ${\tilde c}_{\omega
\mathbb P}$ in Eq.~(\ref{Eq:cuts}) equal to zero.

\begin{table}   
\caption{\label{tab:Fit1fix}Coupling constants for $\pi^0$ and
$\eta$ photoproduction used in Fit I as fixed values. \\}
\begin{tabular}{|c|cc|cc|}
\hline
Reggeon &$\lambda_{\pi^0\gamma}$ &$\lambda_{\eta\gamma}$ &$g^{v}$  &$g^{\mathfrak{t}}$  \\
\hline
$\rho$   & 0.115 & 0.910 & 2.7 & 4.2 \\
$\omega$ & 0.310 & 0.246 & 14.2  &  0.   \\
 \hline
$b_1$    & 0.091 & 0.1   &   0.  & -7.6 \\
 \hline
\end{tabular}
\end{table}

\begin{table*}   
\caption{\label{tab:Fit1par}Parameter values obtained from Fit I and Fit III for
$\pi^0$ and $\eta$ photoproduction.\\}
\begin{tabular}{|c|c|c|c|c|c|c|c|c|c|c|c|c|c|}
\hline
 Solution &Reaction &$c_{\rho\mathbb P}$ &$c_{\omega \mathbb P}$ &$c_{\rho f_2}$ &$c_{\omega f_2}$
&${\tilde c}_{\rho\mathbb P}$ &${\tilde c}_{\omega \mathbb P}$ &${\tilde c}_{\rho f_2}$
&${\tilde c}_{\omega f_2}$ &$d_{\rho\mathbb P}$ &$d_{\omega \mathbb P}$
&$d_{\rho f_2}$ &$d_{\omega f_2}$\\
\hline
I  & $\gamma p \to \pi^0 p$
   &$0.52$    &$-0.06$   &$ 0.72$   &$2.98$    &$0$ &$0$ &$-0.65$   &$0.007$  &$1.07$    &$0.37$    &$0.62$    &$5.02$\\
  & &$\pm0.08$ &$\pm0.01$ &$\pm0.64$ &$\pm0.49$ & -  & -  &$\pm0.26$ &$\pm0.1$ &$\pm0.71$ &$\pm0.14$ &$\pm0.43$ &$\pm0.77$ \\
 \hline
I    &$\gamma p \to \eta p$
   &$-2.27$   &$0.016$  &$5.89$  &$-5.96$&$0$&$0$&$-0.18$     &$0.25$   &$5.5$ &$5.5$ &$2.36$ &$2.36$\\
& &$\pm0.92$&$\pm0.09$&$\pm0.81$&$\pm0.83$& - & - &$\pm0.28$&$\pm0.37$&$\pm2.1$ & -  &$\pm0.19$ & - \\
 \hline
 \hline
III  & $\gamma p \to \pi^0 p$
  &$-0.49$   &$0.23$  &$ 1.08$   &$2.25$&$0$&$0$&$0.24$  &$0.08$   &$0.66$   &$9.9$   &$0.001$   &$4.16$\\
& &$\pm0.09$&$\pm0.01$&$\pm0.84$&$\pm0.32$& - & - &$\pm0.31$&$\pm0.1$&$\pm0.16$&$\pm0.61$&$\pm0.87$&$\pm0.51$ \\
 \hline
III    &$\gamma p \to \eta p$
   &$-2.59$ &$-0.25$ &$6.51$  &$-5.77$&$0$&$0$&$-0.17$   &$-0.13$   &$5.5$ &$5.5$ &$2.49$ &$2.49$\\
& &$\pm0.83$&$\pm0.31$&$\pm0.79$&$\pm0.85$& - & - &$\pm0.33$&$\pm0.39$&$\pm4.4$ & - &$\pm0.18$ & - \\
 \hline
\end{tabular}
\end{table*}

\begin{figure*}
\begin{center}
\resizebox{1.0\textwidth}{!}{\includegraphics{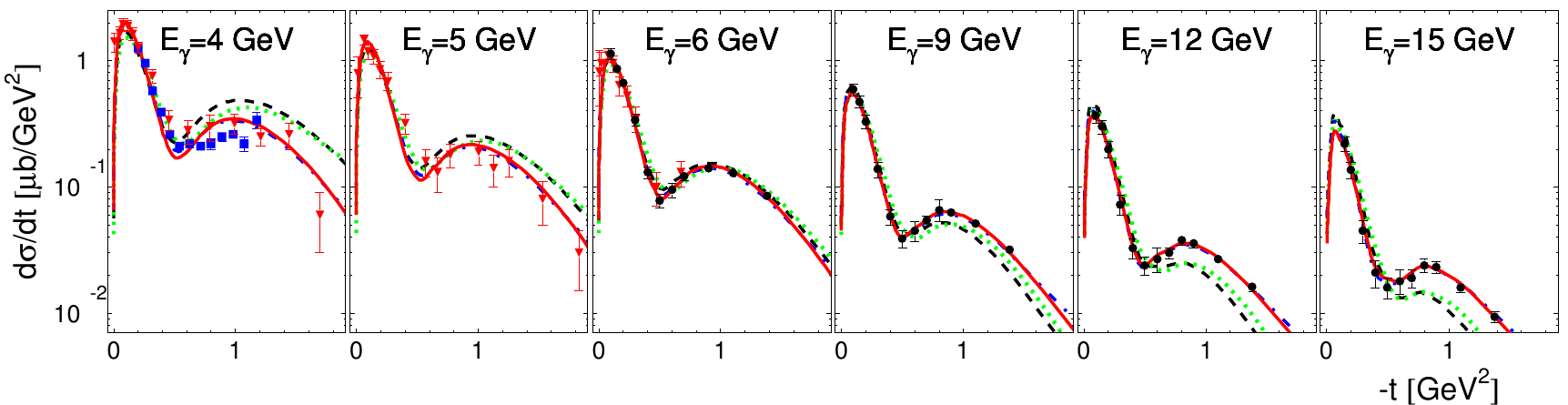}}
\caption{Differential cross sections for $\gamma p \to \pi^0 p$. The
solid red, dashed black, dash-dotted blue (coincide mostly with red
curves), and dotted green lines are our solutions I, II, III, and
IV, respectively. Data are from SLAC~\cite{SLAC:1971} (black
circles) and from DESY:~\cite{DESY:1968} (red triangles)
and~\cite{DESY:1973} (blue squares). } \label{fig:pi0p_dcs}
\end{center}
\end{figure*}
\begin{figure*}
\begin{center}
\resizebox{0.8\textwidth}{!}{\includegraphics{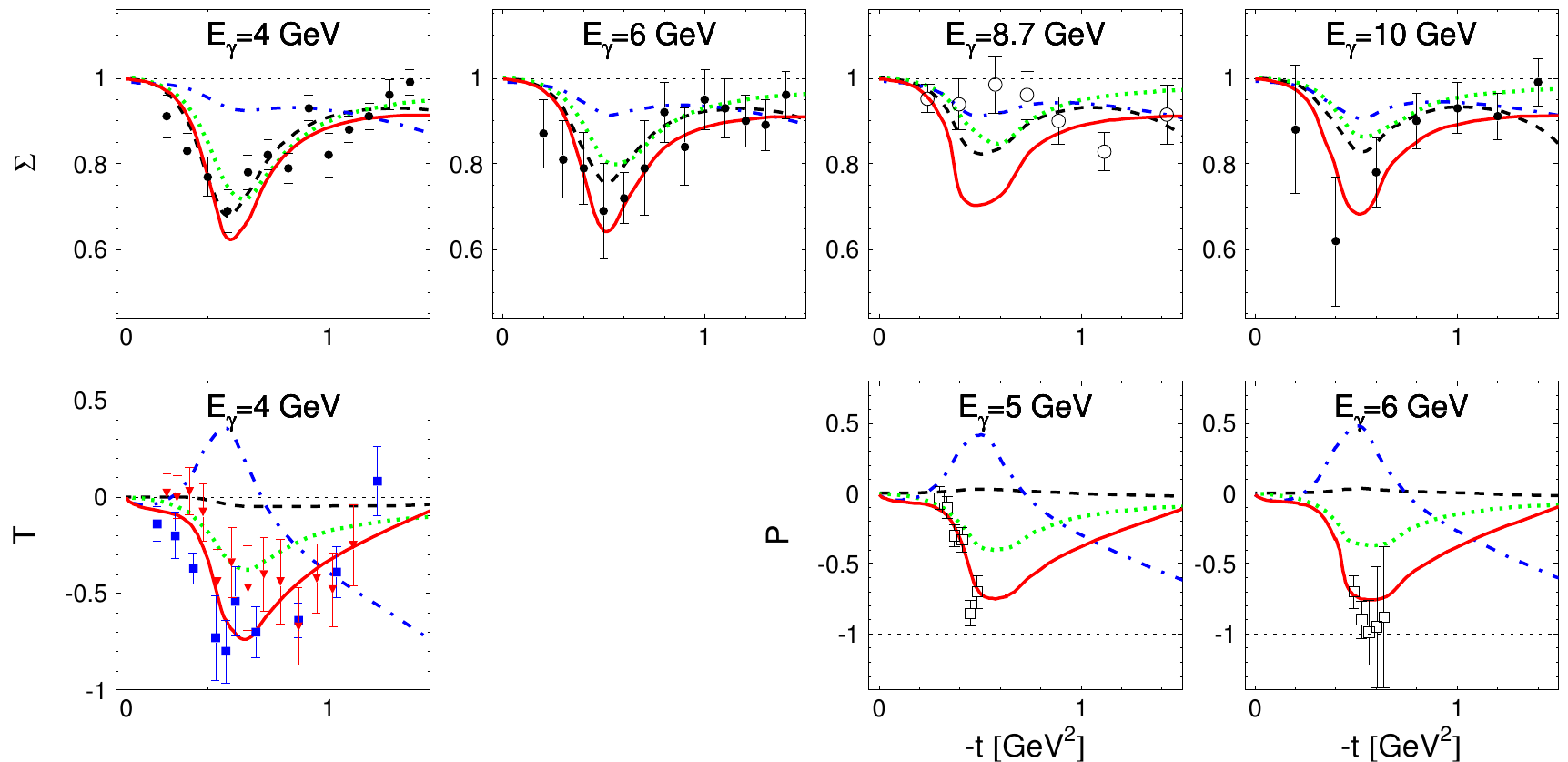}}
\caption{Polarization observables $\Sigma$, $T$, and $P$ for $\gamma
p \to \pi^0 p$. The notation of the lines is the same as in
Fig.~\ref{fig:pi0p_dcs}. Data: SLAC~\cite{SLAC:1971} (black disks),
GlueX-17~\cite{GlueX:2017} (black open circles),
Daresbury~\cite{Dares:1972} (red triangles), DESY~\cite{DESY-T:1973}
(blue full squares), CEA~\cite{CEA:1972} (blue open squares). }
\label{fig:pi0p_stp}
\end{center}
\end{figure*}

\subsection{\boldmath Regge amplitudes and fixed-$t$ dispersion relations}

The formulation of Regge amplitudes as given in the Section III (B)
does not satisfy fixed-$t$ dispersion relations. The reason is
mainly given by the ansatz in Eq.~(\ref{eq:Regge-1}), where the
energy dependence is proportional to $s^{(\alpha(t)-1)}$, violating
crossing symmetry. As an alternative ansatz we also used the
parametrization of Ref.~\cite{Nys:2017} (JPAC model)
\begin{equation}
D_{V,A}
=-\beta_i(t)\;\frac{\pi\,\alpha^\prime_{V,A}(e^{-i\pi\alpha_{V,A}(t)}-1)}
{2\,\mbox{sin}[\pi\alpha_{V,A}(t)]}\;\left({r_i}^{V,A} \nu\right)^{\alpha_{V,A}(t)-1}
\,. \label{eq:Regge-3}
\end{equation}

Here the Mandelstam variable $s$ is replaced by the crossing
variable $\nu$ and the Gamma function in the denominator of
Eq.~(\ref{eq:Regge-1}) is replaced by a more general residue
$\beta_i(t)$, where $i = 1, 2, 3, 4$ is index of the invariant amplitudes.
${r_i}^{V,A}$ are scale parameters of dimension
GeV$^{-1}$. Each exchange, V or A, has its own scale parameter.

In Ref.~\cite{Nys:2017} the following residues for
$V=\rho,\omega$ and $A=b,h$ are given
\begin{eqnarray}
\beta^V_1(t) &=& g^V_1 \,t\, \frac{-\pi \alpha^{\prime V}}{2 }\frac{1}{\Gamma(\alpha^V(t) + 1)} \label{eq:beta_1_V} \,,\\
\beta^V_4(t) &=& g^V_4 \frac{-\pi \alpha^{\prime V}}{2
}\frac{1}{\Gamma(\alpha^V(t))}  \label{eq:beta_4_V} \,,\\
\beta^{\prime A}_2(t) &=& g^A_2 \,t\, \frac{-\pi \alpha^{\prime A}}{2}\frac{1}{\Gamma(\alpha^A(t)+1)} \label{eq:b_gamma_plus_one} \,,
\end{eqnarray}
where the prime in $\beta_2'$ denotes the fact that this is the
$A_2'$ residue, which explains the factor of $t$. The factor $-\pi
\alpha^{\prime}/2$ ensures the correct on-shell couplings. The
functions $1/\Gamma(\alpha+1)$ and $1/\Gamma(\alpha)$ are both equal
to $1$ at the pole $\alpha = 1$, however they differ in the physical
region.

As possible candidates for the $A_3$ amplitude, tensor mesons
$\rho_2$ and $\omega_2$ were suggested in Ref.~\cite{Nys:2017}. The
signature for the tensor mesons is equal to +1, so we use the
following parametrization for the propagator
\begin{equation}
D_T=-\beta_3(t)\;\frac{\pi\,\alpha^\prime_T(e^{-i\pi\alpha_T(t)}+1)}
{2\,\mbox{sin}[\pi\alpha_T(t)]}\;\left({r_i}^{V,A} \nu\right)^{\alpha_T(t)-1}
\label{eq:Regge-4}
\end{equation}
with the residue
\begin{eqnarray}
\beta^T_3(t) &=& g^T_3 \frac{-\pi \alpha^{\prime T}}{2}\frac{1}{\Gamma(\alpha^T(t))}
\label{eq:beta_3_T} \,,
\end{eqnarray}
where a symbol $T$ denotes the tensor meson, $\rho_2$ or $\omega_2$.
Parameters of the trajectories of these mesons are shown in
Table~\ref{tab:traject}. Furthermore, we also assume the same
contributions to $A_3$ from both mesons.

\section{Results}

We have used the Regge cut and JPAC models for a fit to the
available data for $\gamma p \to \pi^0 p$  and $\gamma p \to \eta p$
at $E_\gamma \ge 4$~GeV. The electromagnetic coupling constants for
the $\rho$, $\omega$, and $b_1$ mesons were fixed according to
Table~\ref{tab:Fit1fix}.
The best fit using Regge cuts is called
Solution I.

\begin{table*}      
\caption{\label{tab:chi2}Four solutions using different models and
data sets shown in our analysis.}
\begin{tabular}{|c|c|c|c|c|c|}
\hline
 Solution & Line in Figs. & Model  & Data set & $\chi^2_{red}(\pi)$ & $\chi^2_{red}(\eta)$ \\
\hline
  I & solid red        & Regge cut   & all                         & 1.46 & 1.25  \\
 II & dashed black     & JPAC        & all                         & 5.59 & 2.73  \\
III & dash-dotted blue & Regge cut   & d$\sigma$/dt + GlueX $\Sigma$ & 0.92 & 1.07  \\
 IV & dotted green     & JPAC+$\phi$ & all                         & 4.17 & 1.86  \\
\hline
 \end{tabular}
 \end{table*}
\begin{table*}   
\caption{\label{tab:Fit1par}Partial $\chi^2$ per data points of
$\pi^0$ photoproduction for each observable and each laboratory, for
the solutions I and III.\\}
\begin{tabular}{|c|c|c|c|c|c|c|c|c|c|c|c|}
\hline
 Solution &$d\sigma/dt$  &$d\sigma/dt$ &$d\sigma/dt$  &$\Sigma$
          &$\Sigma$      &$T$          &$T$           &$P$
          &$R_{np}$      &$R_{np}$     &$R_{np}$  \\

          &SLAC~\cite{SLAC:1971}  &DESY~\cite{DESY:1968}
          &DESY~\cite{DESY:1973}  &SLAC~\cite{SLAC:1971}
          &GlueX~\cite{GlueX:2017} &Dares~\cite{Dares:1972}
          &DESY~\cite{DESY-T:1973}  &CEA~\cite{CEA:1972}
          &DESY~\cite{DESY-np:1973}  &CEA~\cite{CEA:1971}
          &Cornell~\cite{Cornell:1973} \\
\hline
 I  & 0.27 &1.56 &14.5 &1.05 &4.27 &1.69 &1.26 &2.94 &3.85 &1.71 &1.19  \\
III & 0.27 &1.36 &9.30 &4.50 &1.05 &25.8 &4.57 &46.2 &7.82 &3.65 &2.82  \\
 \hline
\end{tabular}
\end{table*}
\begin{table*}   
\caption{\label{tab:Fit1par}Partial $\chi^2$ per data points of
$\eta$ photoproduction for each observable and each laboratory, for
the solutions I and III.\\}
\begin{tabular}{|c|c|c|c|c|}
\hline
 Solution &$d\sigma/dt$  &$d\sigma/dt$ &$\Sigma$  &$T$ \\

          &DESY~\cite{DESY:1970}  &WLS~\cite{WLS:1971}
          &GlueX~\cite{GlueX:2017}  &Daresbury~\cite{Dares:1981} \\
\hline
 I  & 1.05 &0.94 &0.44 &2.94 \\
III & 0.98 &0.98 &0.26 &3.80 \\
 \hline
\end{tabular}
\end{table*}

As first step in fits with the JPAC approach, we reproduced
exactly the results from Ref.~\cite{Nys:2017} for the differential cross section
of the $\gamma p \to \eta p$ reaction.  We then added the tensor mesons
$\rho_2$ and $\omega_2$ with electromagnetic couplings fixed to 1
and fitted the model to all available data in $\pi^0$ and $\eta$ production.
This result is called Solution II.

\subsection{\boldmath Results on $\pi^0$ photoproduction}

In the fits we have used the experimental data for the differential
cross sections $d\sigma/dt$ from DESY at $E_{\gamma}=
4$~GeV~\cite{DESY:1973} and $E_{\gamma}= 4, 5,$ and 5.8
GeV~\cite{DESY:1968}, and SLAC~\cite{SLAC:1971} at $E_{\gamma}= 6,
9, 12$, and 15 GeV; the polarized-beam asymmetry $\Sigma$ from
SLAC~\cite{SLAC:1971} at $E_{\gamma}= 4, 6$, and 10 GeV and
GlueX~\cite{GlueX:2017} at $E_{\gamma}= 8.7$~GeV; the target
asymmetry $T$ from Daresbury~\cite{Dares:1972} and
DESY~\cite{DESY-T:1973}, both at $E_{\gamma}= 4$~GeV; the recoil
polarization observable $P$ from CEA~\cite{CEA:1972} at $E_{\gamma}=
4.1 - 6.3$~GeV; the differential cross section ratio of neutrons and
protons, $R_{np}$ for $\pi^0$ photoproduction at $E_{\gamma}= 4$~
GeV~\cite{DESY-np:1973, CEA:1971} and $E_{\gamma}= 4.7$ and 8.2
GeV~\cite{Cornell:1973}.

\begin{figure*}
\begin{center}
\resizebox{0.8\textwidth}{!}{\includegraphics{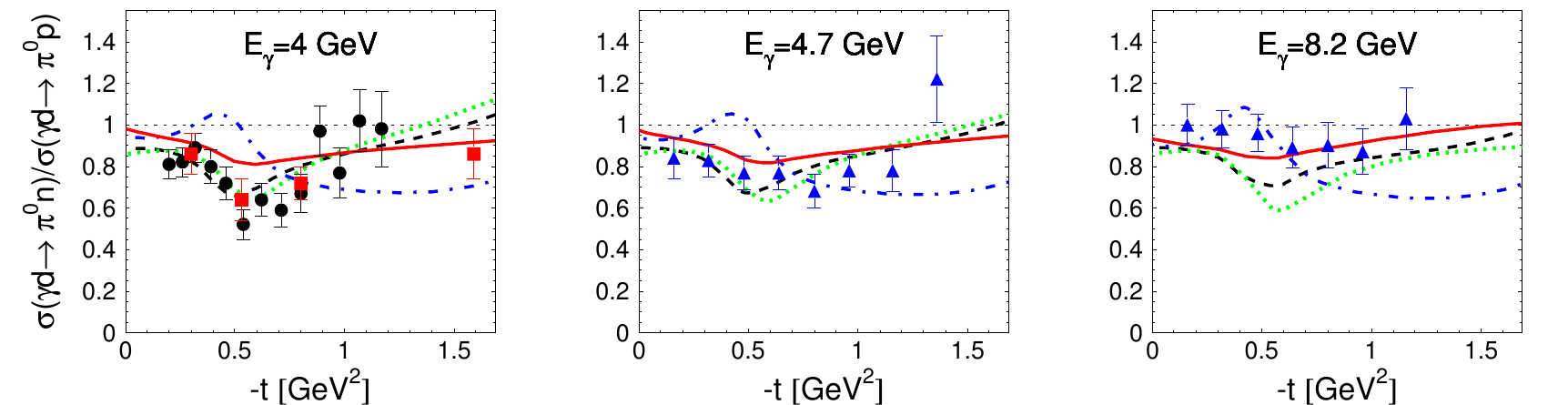}} \caption{Ratio
of differential cross sections for $\pi^0$ photoproduction on
neutrons and protons. The notation of the lines is the same as in
Fig.~\ref{fig:pi0p_dcs}. Data: DESY~\cite{DESY-np:1973} (black
circles), CEA~\cite{CEA:1971} (red squares), and
Cornell~\cite{Cornell:1973} (blue triangles). } \label{fig-np-ratio}
\end{center}
\end{figure*}

The fit results, together with the experimental data, are presented
in Fig.~\ref{fig:pi0p_dcs} for the differential cross sections, in
Fig.~\ref{fig:pi0p_stp} for the polarization observables, and in
Fig.~\ref{fig-np-ratio} for the ratio $R_{np}$. The data for the
recoil polarization observable $P$ are divided in two groups and are
shown on panel $E_{\gamma}= 5$ GeV for $E_{\gamma}= 4.5 - 5.5$~GeV
and on panel $E_{\gamma}$= 6 GeV for $E_{\gamma}= 5.5 - 6.3$~GeV.
The best fit with reduced $\chi^2_{red} =1.46$ using the Regge cut
model is shown by the red lines (Solution I). This solution
describes practically all experimental data except the beam
asymmetry $\Sigma$ at $E_{\gamma}=8.7$~GeV~\cite{GlueX:2017} very
well. The old data from SLAC~\cite{SLAC:1971} for $\Sigma$ at
$E_{\gamma}= 6$ and 10 GeV show a clear dip at $t= -0.5$~GeV$^2$.
Surprisingly, such a structure is missing for the intermediate
energy of 8.7 GeV in the new GlueX data~\cite{GlueX:2017}.
Therefore, we also performed an alternative fit using the Regge cut
model without the old polarization data and obtained the Solution
III with $\chi^2_{red}= 0.92$, which is shown in
Figs.~\ref{fig:pi0p_dcs},~\ref{fig:pi0p_stp},~\ref{fig-np-ratio} by
the dash-dotted blue line. This solution can describe the GlueX data
quite well, but it is absolutely wrong for $T$ and $P$ and also
underestimates the old data for $\Sigma$. Therefore, we conclude,
that a strong energy dependence of the beam asymmetry between 6 and
10 GeV, as suggested by the GlueX data, cannot be described within
our model without adding additional dynamics.
There is also some disagreement between the data and the Solution I
for the differential cross sections at $E_{\gamma}= 4$~GeV, see
Figs.~\ref{fig:pi0p_dcs} and~\ref{fig-np-ratio}. This energy
corresponds to the center-of-mass energy $W = 2.9$~GeV, that is
close to the resonance region. Probably, tails from the resonance
contributions still show up in this energy region for $\pi^0$
photoproduction and should be take into account.

The central values of the fit parameters for the Solution I and III
are shown in Table~V together with associated uncertainties.
Parameters without errors were fixed in the fits. The coefficients
${\tilde c}_{\rho\mathbb P}$ and ${\tilde c}_{\omega \mathbb P}$ are
zero because the corresponding terms for the $\rho\mathbb P$ and
$\omega\mathbb P$ cuts do not contribute to the $A_3$ amplitude, see
Table III. There are also two parameters for the $\gamma p \to \eta
p$ reaction that were fixed by empirical constraints: $d_{\omega
\mathbb P}$ = $d_{\rho\mathbb P}$ and $d_{\omega f_2}$ = $d_{\rho
f_2}$.

The best fit with the JPAC model has $\chi^2_{red}= 5.59$ (Solution
II), see black dashed lines in
Figs.~\ref{fig:pi0p_dcs},~\ref{fig:pi0p_stp}. It describes well the
shape of the differential cross sections but has the wrong energy
dependence after the dip location, $-t > 0.4$~GeV$^2$. Similar to
the Regge cut solution, it does not describe the new GlueX data for
$\Sigma$. Furthermore, the existing data on the polarization
observables $T$ and $P$  cannot be described. The inclusion of the
exotic tensor mesons $\rho_2$ and $\omega_2$ did not improve our
fits and we did not consider them in our four solutions.

We then investigated the possibility of improving the fit by
including the $\phi$ meson in the JPAC model even though small
couplings to the nucleon can be expected as discussed above. The
electromagnetic coupling constants $\lambda_{\phi\pi^0\gamma}$=
0.018 and $\lambda_{\phi\eta\gamma}$= 0.38 are obtained from the
corresponding widths $\Gamma_{\phi\rightarrow \pi^0\gamma} = 5.4\
\mathrm{keV}$ and $\Gamma_{\phi\rightarrow \eta\gamma} = 55.84\
\mathrm{keV}$~\cite{PDG2016} using Eq.~(\ref{Eq:Gamma}). This solution IV is
shown in Figs.~\ref{fig:pi0p_dcs},~\ref{fig:pi0p_stp},~\ref{fig-np-ratio}
by the green dotted lines. We did not use $\rho_2$
and $\omega_2$ for this fit because of their negligible
contributions. Indeed, Solution~IV describes the polarization
observables $T$ and $P$ significantly better than Solution~II. The
hadronic vector $g^v=-4.3$ and tensor $g^t=-0.08$ coupling constants
for $\phi$ meson were obtained from this fit which we consider as
reasonable. A comparison of $\chi^2_{red}$ for the different solutions is shown in Table~VI.

Table VII gives partial $\chi^2$ divided by the number of the data
points for each observable and each laboratory, for the solutions I
and III.

\subsection{\boldmath Results on $\eta$ photoproduction}

\begin{figure*}
\begin{center}
\resizebox{0.8\textwidth}{!}{\includegraphics{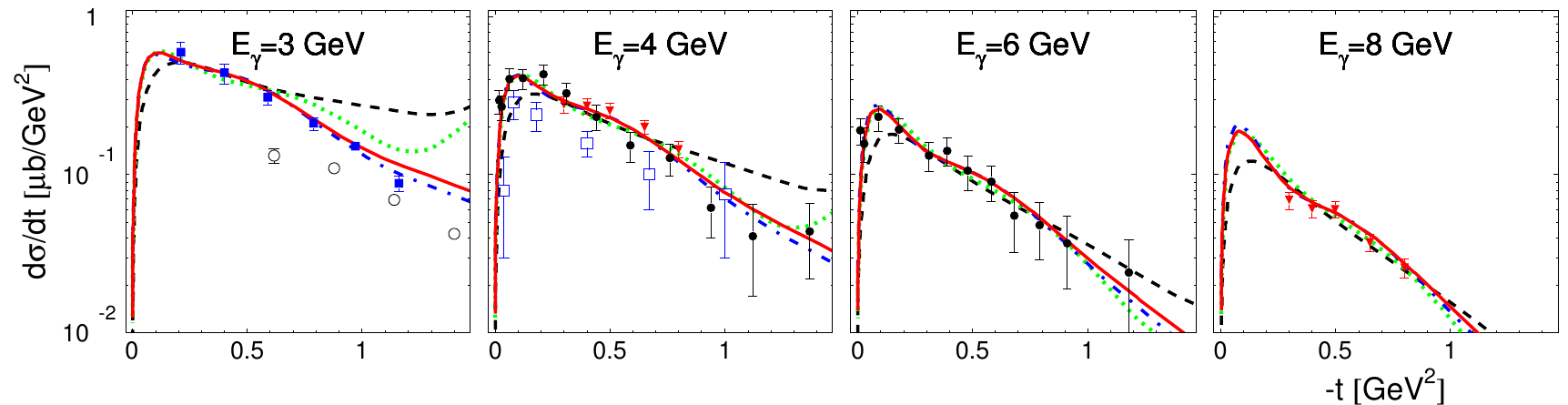}}
\caption{Differential cross sections for $\gamma p \to \eta p$. The
notation of the lines is the same as in Fig.~\ref{fig:pi0p_dcs}.
Data: DESY~\cite{DESY:1970} (black disks), WLS~\cite{WLS:1971} (red
triangles), Daresbury~\cite{Dares:1976} (blue full squares),
CLAS~\cite{CLAS:2009} (black open circles), and CEA~\cite{CEA:1968}
(blue open squares). } \label{fig:eta_dcs}
\end{center}
\end{figure*}
%
\begin{figure*}
\begin{center}
\resizebox{0.8\textwidth}{!}{\includegraphics{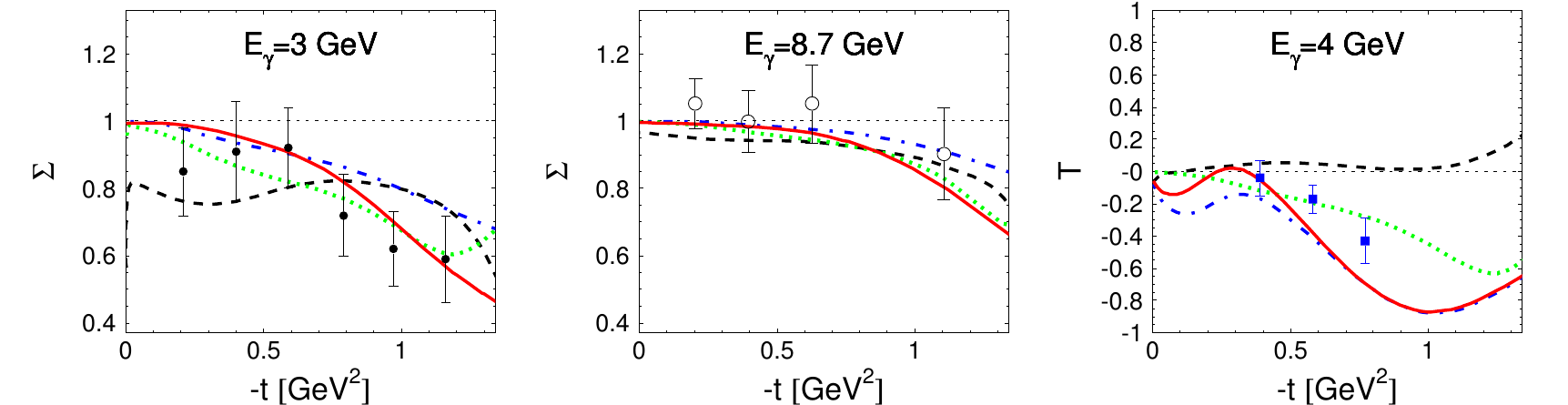}}
\caption{Polarization observables $\Sigma$ and $T$ for $\gamma p \to
\eta p$. The notation of the lines is the same as in
Fig.~\ref{fig:pi0p_dcs}. Data: GlueX~\cite{GlueX:2017} (black open
circles) and Daresbury:~\cite{Dares:1976} (black disks)
and~\cite{Dares:1981} (blue squares). } \label{fig:eta_st}
\end{center}
\end{figure*}
%
The data set for the $\gamma p \to \eta p$ reaction at high energies
is more limited than for $\pi^0$ photoproduction. For the fit, we
have used the experimental data of the differential cross sections
$d\sigma/dt$ from DESY~\cite{DESY:1970} at $E_{\gamma}$= 4 and 6 GeV
and WLS~\cite{WLS:1971} at $E_{\gamma}$= 4 and 8 GeV; for the
polarized-beam asymmetry $\Sigma$ from GlueX~\cite{GlueX:2017} at
$E_{\gamma}= 8.7$~GeV; and for the target asymmetry $T$ from
Daresbury~\cite{Dares:1981}.

Our fit results for the differential cross sections are presented in
Fig.~\ref{fig:eta_dcs} and for the polarization observables $\Sigma$
and $T$ in Fig.~\ref{fig:eta_st}. The data for $d\sigma/dt$ and
$\Sigma$ at $E_{\gamma}= 3$~GeV were not included in the fit,
because these are very close to the resonance region. However, the
predictions of all our solutions can reproduce also these data quite
well. Presumably, the influence of the resonances for $\eta$
photoproduction is already negligible at these energies.
Our extrapolation of the differential cross section to $E_{\gamma}=
3$~GeV is in good agreement with Ref.~\cite{Sibirtsev3}.

The best fit with $\chi^2_{red}$= 1.25 using the Regge cut model is
shown by the solid red line (Solution I). This solution well
describes all experimental data including the beam asymmetry
$\Sigma$ at $E_{\gamma}=8.7$ GeV~\cite{GlueX:2017}. The alternative
fit without data for $T$, Solution III, also gives a good prediction
for this observable.

Table VIII gives partial $\chi^2$ divided by the number of the data
points for each observable and each laboratory, similar as in Table
VII, but for $\eta$ photoproduction.

The fit with the JPAC model has a $\chi^2_{red}=2.73$ (Solution II),
see dashed black lines in Figs.~\ref{fig:eta_dcs},\ref{fig:eta_st}.
Similar as for $\pi^0$ photoproduction, it well describes the
differential cross section and $\Sigma$, but contradicts the data
for $T$. As in case of $\pi^0$ production, the inclusion of the
$\phi$ meson (Solution IV), improves the description significantly
at low $t$. However, a main drawback of Solution IV is a large
overestimation of the total cross section at energies $E_{\gamma} >
2$~GeV. Therefore, this solution can not be used as a non-resonant
background for partial wave analyses in the resonance region.

\subsection{\boldmath Further results for high energies}

From high-energy approximations of the observables the following
relation between the target and recoil polarization to the photon
beam asymmetry can be derived in a model independent way (see appendix):
\begin{equation}
|P-T| \le 1 -\Sigma
\end{equation}
\begin{figure}
\begin{center}
\resizebox{0.49\textwidth}{!}{\includegraphics{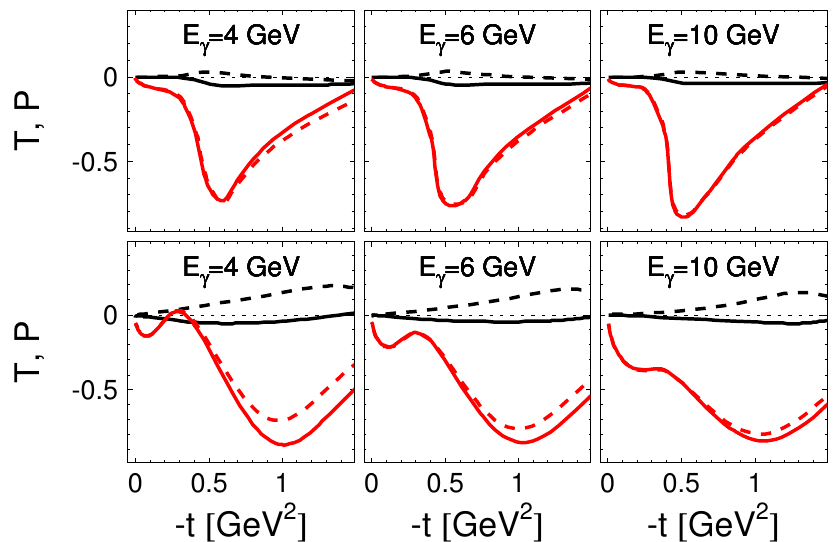}}
\caption{Comparison of the polarization observables $T$ and $P$ at
different photon beam energies for $\gamma p \to \pi^0 p$ (top
panels) and for $\gamma p \to \eta p$ (bottom panels). The solid red
and black lines are our solutions I and II for the target
polarization $T$ and the dashed red and black lines for the recoil
polarization $P$, respectively. } \label{fig:eta_tp}
\end{center}
\end{figure}
As the beam asymmetry $\Sigma$ is almost unity, except in the
neighborhood of the dip near $t= -0.5$~GeV$^2$, the polarization
observables $T$ and $P$ should be almost equal. Any difference
between $T$ and $P$ should be due to an interference between the
$A_2'$ and $A_3$ amplitudes at high energies, see Eqs.~(C3),(C4) in
Appendix C. A comparison between $T$ and $P$ for the Solutions I and
II is shown in Fig.~\ref{fig:eta_tp}. The Solution I for $\pi^0$
photoproduction verifies well this prediction. There is some visible
difference between $T$ and $P$ for $\eta$ photoproduction, but in
this case no $P$ data were included in the fit.


\section{Summary and conclusions}

Photoproduction $\pi^0$ and $\eta$ mesons on the nucleon at photon
beam energies above 4 GeV was investigated within two different
Regge model approaches. The models include $t$-channel exchange of
vector ($\rho$ and $\omega$) and axial vector ($b_1$ and $h_1$)
mesons. Moreover, Regge cuts of $\rho{\mathbb P}$, $\rho f_2$,
$\omega{\mathbb P}$, and $\omega f_2$ are used. Both models can
describe differential cross sections and photon beam asymmetries
$\Sigma$ very well, except for a possible strong energy dependence
of $\Sigma$ in $\gamma p \to \pi^0 p$ between 6 and 10 GeV as
suggested by recent GlueX data. Within our approach we can not find
a solution that can simultaneously describe both the old
polarization data and the new GlueX data.

The crossing-odd amplitude $A_3$ gets no contributions from dominant
$t$-channel vector meson exchange terms. We found possible
contributions from tensor meson exchanges and also from Regge cuts.
All of them turn out to be rather small. The effect could be worked
out in the difference between target and recoil polarizations, but
from existing data in $\pi^0$ photoproduction no evidence can be
seen.

Finally, with the present database only the Regge cut model
(Solution I) is able to describe all other available polarization
observables as well. However, since most data go back to the late
1960s and early 1970s, and on the other hand new data are in
progress, a reliable conclusion can not yet be drawn. For our
applications in forthcoming baryon resonance analyses from
pseudoscalar meson photoproduction data, we currently favor an
extrapolation of Solution I to lower energies as a good description
for the non-resonant background.
\clearpage

\begin{acknowledgments}
We would like to thank V.~Mathieu, J.~Nys and M.~Vanderhaeghen for
very fruitful discussions. This work was supported by the Deutsche
Forschungsgemeinschaft (SFB 1044).
\end{acknowledgments}

\begin{appendix}

\section{Observables in terms of CGLN amplitudes}

Here the polarization observables involving beam and target
polarization are expressed by helicity amplitudes in the notation of
Barker~\cite{Barker} and Walker~\cite{Walker}. A phase space factor
$|{\bold q}|/|{\bold k}|$ has been omitted in all expressions. The
differential cross section is given by $\sigma_0$ and the spin
observables $\check{O}_i$ are obtained from the spin asymmetries
$A_i$ by $\hat{O}_i=A_i\,\sigma_0$:
\newcommand{\fpf}[2]{{F}_{#1}^{*}{F}_{#2}}
\begin{eqnarray}
\begin{split}
\sigma_{0} = &\mbox{Re}\,[\fpf{1}{1} + \fpf{2}{2} + \sin^{2}\theta\,(\fpf{3}{3}/2 + \fpf{4}{4}/2 \\
             &+ \fpf{2}{3} + \fpf{1}{4} + \cos\theta\,\fpf{3}{4}) - 2\cos\theta\,\fpf{1}{2}]\,,  \\
\check{\Sigma} = &- \sin^{2}\theta\;\mbox{Re}\,[(\fpf{3}{3} +\fpf{4}{4})/2 + \fpf{2}{3} + \fpf{1}{4} \\
                 &+ \cos\theta\,\fpf{3}{4}] \,, \\
\check{T}      = &\sin\theta\;\mbox{Im}\,[\fpf{1}{3} - \fpf{2}{4} + \cos\theta\,(\fpf{1}{4} - \fpf{2}{3})\\
                 &- \sin^{2}\theta\,\fpf{3}{4}] \,, \nonumber\\
\check{P}      = &- \sin\theta\;\mbox{Im}\,[2\fpf{1}{2} + \fpf{1}{3} - \fpf{2}{4} \\
                 &- \cos\theta\,(\fpf{2}{3} -\fpf{1}{4}) - \sin^{2}\theta\,\fpf{3}{4}] \,, \nonumber\\
\check{E}      = &\mbox{Re}\,[\fpf{1}{1} + \fpf{2}{2} - 2\cos\theta\,\fpf{1}{2} \\
                 &+ \sin^{2}\theta\,(\fpf{2}{3} + \fpf{1}{4})] \,, \nonumber\\
\check{F}      = &\sin\theta\;\mbox{Re}\,[\fpf{1}{3} - \fpf{2}{4} - \cos\theta\,(\fpf{2}{3} - \fpf{1}{4})]\,,\\
\check{G}      = &\sin^{2}\theta\;\mbox{Im}\,[\fpf{2}{3} + \fpf{1}{4}] \,, \nonumber\\
\check{H}      = & \sin\theta\;\mbox{Im}\,[2\fpf{1}{2} + \fpf{1}{3} - \fpf{2}{4} \\
                 &+ \cos\theta\,(\fpf{1}{4} - \fpf{2}{3})] \,. \nonumber\\
\end{split}
\end{eqnarray}

\section{CGLN amplitudes in terms of invariant amplitudes}
\label{app:FtoA}
The CGLN amplitudes are obtained from the
invariant amplitudes $A_i$ by the following equations
\cite{Dennery:1961, Ber:1967}:
\begin{eqnarray}
\begin{split}
{F}_1 =& \frac{W-M_N}{8\pi\,W}\,\sqrt{(E_i+M_N)(E_f+M_N)}\big[ A_1  \\
       &+(W-M_N)\,A_4 - \frac{2M_N\nu_B}{W-M_N}\,(A_3-A_4)\big]\,,\nonumber \\
{F}_2 =& \frac{W+M_N}{8\pi\,W}\,|{\bold q}|\,\sqrt{\frac{E_i-M_N}{E_f+M_N}}\big[-A_1 + (W+M_N)\,A_4\\
       &- \frac{2M_N\nu_B}{W+M_N}\,(A_3-A_4)\big]\,, \nonumber \\
{F}_3 =& \frac{W+M_N}{8\pi\,W}\,|{\bold q}|\,\sqrt{(E_i-M_N)(E_f+M_N)}\big[(W-M_N)\,A_2 \\
       &+ A_3-A_4\big]\,,  \nonumber \\
{F}_4 =& \frac{W-M_N}{8\pi\,W}\,{\bold q}^2\,\sqrt{\frac{E_i+M_N}{E_f+M_N}} \big[-(W+M_N)\,A_2 \\
       &+A_3 - A_4\big]\,,
\end{split}
\end{eqnarray}

with $\nu_B=(t-\mu^2)/(4M_N)$.

\section{Observables in terms of invariant amplitudes}

For high energies, the polarization observables can conveniently be
described in terms of invariant amplitudes. Here we follow
Ref.~\cite{Mathieu:2015} and derive the expressions at leading order
in the energy squared:
\begin{eqnarray}
 \frac{d\sigma}{dt} &\approx&  \frac{1}{32\pi} \left[|A_1|^2+|A_2'|^2-t|A_3|^2-t|A_4|^2\right]\,, \\
\Sigma \frac{d\sigma}{dt} &\approx&  \frac{1}{32\pi} \left[|A_1|^2-|A_2'|^2+t|A_3|^2-t|A_4|^2\right]\,, \\
T \frac{d\sigma}{dt} &\approx&  \frac{1}{16\pi} \sqrt{- t}\; \text{Im}\left[A_1\,A_4^*-A_2'\,A_3^* \right]\,,\\
P \frac{d\sigma}{dt} &\approx&  \frac{1}{16\pi} \sqrt{- t}\; \text{Im}\left[A_1\,A_4^*+A_2'\,A_3^*
\right]\,.
\end{eqnarray}
From these relations, a restriction for the difference between
target and recoil polarization can be found
\begin{equation}
|P-T| \le 1 -\Sigma\,.
\end{equation}

\end{appendix}

\newpage

\end{document}